\documentclass[12pt]{article}
\usepackage{amssymb}
\usepackage{eucal}
\begin{document}
\newcommand{\ben}{\begin{eqnarray}}
\newcommand{\een}{\end{eqnarray}}
\newcommand{\la}{\label}
\vskip 1truecm
\centerline{\huge Applications of Lobachevsky Geometry }

\centerline{{\huge to the Relativistic Two-Body
Problem}{\Large\footnote{Talk given at the International seminar:
{\em Application and Development of Lobachevsky Ideas in Modern
Physics}, dedicated to 75-th birthday of Professor
N.~A.~Chernikov, Dubna, 25-27 February 2004}}}
\vskip 1truecm
\centerline{\Large Plamen P.~Fiziev \footnote{ E-mail:\,\,
fiziev@phys.uni-sofia.bg}}
\vskip 1truecm
\centerline{\em Department of Theoretical Physics, Sofia
University,}

\centerline{\em 5 James Bourchier Boulevard, BG-1164, Sofia,
Bulgaria}
\vskip 1truecm
\begin{abstract}
In this talk we consider the geometrical basis for the reduction
of the relativistic 2-body problem, much like the non-relativistic
one, to describing the motion of an effective particle in an
external field. It is shown that this possibility is deeply
related with the Lobachevsky geometry. The concept of relativistic
reduced mass and effective relativistic particle is discussed
using this geometry. Different recent examples for application of
relativistic effective particle are described in short.
\end{abstract}

\sloppy

\section{Introduction}

The classical (non-quantum) relativistic two-body problem has a
long history. Although at present many different approaches to
this problem exist, we still do not have a satisfactory solution
in many cases of different interactions between the two massive
point particles. Much has been done in the case of relativistic
interaction of two electrically charged particles \cite{Fock, LL,
WF49, K72}. An essential progress is reached in the case of
two-particle gravitational interaction \cite{Fock, LL, K72}, in
the case of two massive relativistic particles with string
interaction between them \cite{BN} and in the case of areal
interaction \cite{ChSh}.

There exist several basic obstacles to general solution of the
problem: the existence of many different possible choices of the
time variable, the absence of a unique choice of the
center-of-mass position variables, the complications, which appear
in the attempts to introduce a proper phase-space of relativistic
two-particles and during the reduction of their degrees of
freedom, the correct inclusion of relativistic retardation and
corresponding non-locality of the interactions, the nonlinear
character of typical relativistic interactions between particles,
like gravitational interaction and string one, the problem how to
take account of the internal spin both of the interacting
particles and of the carrier of the interaction, etc.

In the present talk we will concentrate on one of the possible
approaches to the relativistic two-particle problem, which is
based on the introduction of an effective particle -- {\em the
relativistic reduced mass}. This approach is analogous to the
non-relativistic one and was introduced in a heuristic way at
first in the quantum relativistic problems by I. Todorov
\cite{Todorov}.

A new derivation of the basic notions of the relativistic
effective one-particle approach to the relativistic two-particle
problem and some applications were given in \cite{FT01}. It turns
out that this derivation lies essentially on the Lobachevsky
geometry in the velocity space of the relativistic particles. In
the present talk we will give a more detailed derivation of the
notion of relativistic effective particle, based on this geometry.
We also review in short some of the recent developments and
applications of this idea \cite{developments}.

\section{The Kinematics of One Relativistic Massive Particle
and Lobachevsky Geometry}

In this section we remind the reader the basic notions of the
simple kinematics of one {\em free} relativistic particle with
mass $m>0$ and its relation with Lobachevsky geometry in the
velocity space .

In this section we choose the laboratory frame system $K$ and and
laboratory time $t$. Further on we use the units in which $c=1$.

\subsection{The Kinematics of One Free Relativistic Massive Particle}
The 3D position of a point particle is described by its
radius-vector $\vec{x}$ and 3D-velocity $\vec{v}$. We use a
standard notations $\{x^0,\vec{x}\}$ ($x^0=t$) for the Cartesian
coordinates in the 4D pseudo-Euclidean space-time
$\mathbb{E}^{(1,3)}$ and a standard action ${\cal A}_m=-m\int
ds=-m\int\sqrt{1-\vec{v}\,{}^2} dt$. The 4D canonical momentum $p$
has components $\{p_0,\vec{p}\}$: \ben
p_0={m\over\sqrt{1-\vec{v}\,{}^2}},\,\,\,
\vec{p}={{m\vec{v}}\over\sqrt{1-\vec{v}\,{}^2}}. \la{p_comp}\een
According to our convention the 4D velocity is $u=p/\sqrt{-p^2}$.
As a consequences one obtains \ben
p=mu,\,\,\,\,\,p^2=\vec{p}\,{}^2-p_0^2\stackrel{w}{=}-m^2,
\,\,\,\,\,u^2=\vec{u}\,{}^2-u_0^2\stackrel{w}{=}-1.\la{hiperb}\een
Here the simbol "$\stackrel{w}{=}$" denotes a "weak" equality,
i.e. a constraint in Dirac's sense \cite{D49}. These relations
define the relativistic 4D "hyperboloid of velocities". In
addition we have for the 3D velocities the useful relations \ben
\vec{u}={{\vec{v}}\over\sqrt{1-\vec{v}\,{}^2}},\,\,\,\,\,
\vec{v}={{\vec{u}}\over\sqrt{1+\vec{u}\,{}^2}},\,\,\,\,\,
\left(1-\vec{v}\,{}^2\right)\left(1+\vec{u}\,{}^2\right)=1.
\la{vu}\een

Let $K^U$ is an inertial frame which moves with respect to the
laboratory frame $K$ with a 4D velocity $U$. Then the velocity of
the mass $m$  with respect the system $K^U$ is \ben
u^U=\Lambda_U^{-1}u,\la{uU}\een where $\Lambda_U\in SO(1,3)$ is a
pure Lorentz transformation that carries the quantities from frame
$K$ into  the frame $K^U$. The conditions $\Lambda^\mu{}_\nu
U^\nu=\delta^\mu_0$ and positive definiteness determine the
(symmetric) Lorentzian matrix $\Lambda$ uniquely:
\begin{eqnarray}
 \Lambda=\pmatrix{U^0 &-U_j\cr
                 -U^i& \delta^i_j+{{U^iU_j}\over{1+U^0}}\cr
                 }.
\label{LambdaU}
 \end{eqnarray}
Here and further on the Latin indexes take values $i,j,...=1,2,3;$
and the Greek ones -- the values $\alpha,\beta,...=0,1,2,3;$ .

 The formula (\ref{uU}) is analogous to the corresponding Galileo
 formula for superposition of velocities in non-relativistic mechanics:
 \ben \vec{v} \to \vec{v}\,{}^V = g_V^{-1}(\vec{v})=
 \vec{v}+\vec{V}.\la{Galileo}\een It describes a Galileo's translation
 in the velocity space $g_V^{-1}\in
 G_{10}$, i.e. the transition from laboratory frame $K$
 to a moving inertial frame $K^V$.
 Here $G_{10}$ is the Galileo's group.

In a transparent form the relation (\ref{uU}) gives \ben
u^U_0=U_0u_0-\vec{U}\!\cdot\vec{u},\hskip 5.8truecm \nonumber \\
\vec{u}\,{}^U\!=\!U_0\,\vec{u}\!-\!u_0\,\vec{U}\!+\!
{{\vec{U}\times\left(\vec{U}\times\vec{u}\right)}\over{1\!+\!U_0}}
\!=\!U_0\,\vec{u}_{{}_{||}}\!-\!u_0\,\vec{U}\!+\!\vec{u}_\perp.\la{uUT}
\een Here $$\vec{u}_{{}_{||}}={\cal P}_U\vec{u},$$
$$\vec{u}_\perp={\cal P}_U^\perp\vec{u},$$ where $${\cal
P}_U=\vec{U}\otimes\vec{U}/U^2$$ and  $${\cal
P}_U^\perp={Id}_3-\vec{U}\otimes\vec{U}/U^2$$ are the
corresponding projector operators, and $Id_3$ is the 3D identity
operator.

Using the 3D velocity $\vec{V}$ we obtain for the 4D velocity
$U=\left\{ {1\over{\sqrt{1-\vec{V}\,{}^2}}},
{{\vec{V}}\over{\sqrt{1-\vec{V}\,{}^2}}}\right\}$ and the standard
form of Lorenz transformation: \ben
u_0^U={{u_0-\vec{V}\cdot\vec{u}}\over{\sqrt{1-\vec{V}\,{}^2}}},
\,\,\,\,\,\,\vec{u}\,{}^U={{\vec{u}_{{}_{||}}-
u_0\vec{V}}\over{\sqrt{1-\vec{V}\,{}^2}}}+\vec{u}_\perp.
\la{Lorenz}\een

For the transformation of the 4D momentum one obtains an analogous
relations: \ben p^U=\Lambda_U^{-1}p,\la{pU}\een \ben
p^U_0\!=\!U_0\, p_0\!-\!\vec{U}\!\cdot\vec{p},\,\,\,\,\,
\vec{p}\,{}^U\!=\!U_0\,\vec{p}\!-\!p_0\,\vec{U}\!+\!
{{\vec{U}\times\left(\vec{U}\times\vec{p}\right)}\over{1\!+\!U_0}}
 \!=\!U_0\,\vec{p}_{{}_{||}}\!-\!p_0\,\vec{U}\!+\!\vec{p}_\perp.\la{pUT}
\een

As we see, we have a point transformation in the momentum space,
which can be easily extended to a canonical transformation on the
whole phase space $\mathbb{M}^{(8)}_{p,x} $ : $\{p,x\} \to
\{p,x\}$ of the type $ x\,dp + p^U dx^U = dF(p,x^U)$ with a
generating function \ben F(p,x^U)=
\vec{p}_\perp\cdot\vec{x}\,{}^U+
U_0\,\vec{p}_{{}_{||}}\cdot\vec{x}\,{}^U-
p_0\,\vec{U}\cdot\vec{x}\,{}^U -
\left(U_0\,p_0-\vec{U}\cdot\vec{p}\right)(x^0)^U.\la{F} \een

Using the formulas $x^0=-\partial_{p_0}F$ and $\vec{x}=
\partial_{\vec{p}}F$ one easily
obtains the standard Lorenz transformations:

\ben x^0=U_0(x^0)^U+\vec{U}\cdot\vec{x}\,{}^U=
{{(x^0)^U+\vec{V}\cdot\vec{x}\,{}^U}\over{\sqrt{1-\vec{V}\,{}^2}}},\nonumber
\\
\vec{x}= (\vec{x}_\perp)^U +
U_0(\vec{x}_{{}_{||}})^U+(x_0)^U\vec{U}=(\vec{x}_\perp)^U +
{{(\vec{x}_{{}_{||}})^U+(x^0)^U\vec{V}}\over{\sqrt{1-\vec{V}\,{}^2}}};\nonumber\\
(x^0)^U=
{{(x^0)-\vec{V}\cdot\vec{x}}\over{\sqrt{1-\vec{V}\,{}^2}}},\nonumber
\\
\vec{x}=\vec{x}_\perp +
{{(\vec{x}_{{}_{||}})-x^0\vec{V}}\over{\sqrt{1-\vec{V}\,{}^2}}}.
\la{Lorenz_x} \een

\subsection{The Lobachevsky Geometry in the Velocity Space}

Following Fock \cite{Fock}, let us consider an infinitesimal
change of the velocity $\vec{U}=\vec{u}+d\vec{u}$. Then from the
representations $\vec{u}\,{}^U=\vec{u}+d\vec{u}\,{}^U$,
$d\vec{u}=d\vec{u}_\perp+d\vec{u}_{{}_{||}}/u_0$, where
$\vec{u}_{{}_{||}}=(\vec{u}\cdot d\vec{u}/\vec{u}^2)\vec{u}$,
$d\vec{u}_\perp=d\vec{u}-d\vec{u}_{{}_{\\}}$ one easily obtains
the metric in the velocity space in the form: \ben
d\sigma^2=(d\vec{u}\,{}^U)^2=g_{ij}(d\vec{u}\,{}^U)^i(d\vec{u}\,{}^U)^j,
\la{metric_u}\een where the metric tensor is given by the $3\times
3$ matrix \ben
g(\vec{u})=Id_3-{{\vec{u}\otimes\vec{u}}\over{1+\vec{u}\,{}^2}}.
\la{gij}\een In spherical coordinates
$\vec{u}=\{\rho_u\cos\theta\cos\varphi,\rho_u\cos\theta\sin\varphi,
\rho_u\sin\theta\}$ the metric in the velocity space reads:
 \ben
d\sigma^2={{(d\rho_u)^2}\over{1+\rho_u^2}}+\rho_u^2\left((d\theta)^2+\sin^2\theta
(d\varphi)^2 \right). \la{metric_u_sph}\een This formula shows
explicitly that the $\vec{u}$-space is a Riemannian one with a
constant negative curvature $-1$, i.e. a Lobachevsky space. It is
just the hyperboloid $u^2=-1$  in the 4D pseudo-Euclidean
space-time $\mathbb{E}^{(1,3)}$ and its metric (\ref{metric_u}) is
just the restriction:
$$d\sigma^2=\Big((d\vec{u})^2-(du_0)^2\Big)_{u_0=\pm\sqrt{1+\vec{u}\,{}^2}}.$$

The length $\sigma(u,U)$ of a geodesic line, which connects two
points $u$ and $U$ on this hyperboloid, is defined by the
formulas: \ben
\sinh(\sigma(u,U))=\sqrt{\left(U_0\,\vec{u}-u_0\,\vec{U}\right)^2
-\left(\vec{U}\times\vec{u}\right)^2,}\nonumber
\\ \cosh(\sigma(u,U))=U_0\,u_0-\vec{U}\!\cdot\vec{u}
\,\,.\hskip 3.1truecm \la{length}\een

\section{Effective Particles in the Two Particle Problem}

\subsection{The Non-Relativistic Two Particle Problem}

It is well known from the textbooks on classical mechanics, that
one can reduce the two-particle problem to one-particle one,
introducing two effective particles: the center of mass (CM) and
the reduced mass $m$. In this section we give some nonstandard way
of introducing a reduced mass in the non-relativistic case. Our
specific approach is based on the momentum space considerations
and allows a proper relativistic generalization.

In the non-relativistic mechanics the CM is an effective particle
with mass $M:=m_1+m_2$, where $m_{1,2}$ are the rest masses of the
physical point particles at 3D positions with radius-vectors
$\vec{r}_{1,2}$ and 3D velocities $\vec{v}_{1,2}$. It is supposed
that the particles are moving in vacuum, i.e., without influence
of any external forces, but they can interact between themselves.
The position vector of the CM and its velocity in the laboratory
frame $K$ are \ben
\vec{R}:={{m_1}\over{M}}\vec{r}_1+{{m_2}\over{M}}\vec{r}_2,
\,\,\,\,\,\vec{V}={{m_1}\over{M}}\vec{v}_1+{{m_2}\over{M}}\vec{v}_2.
\la{RV}\een

By definition the CM-frame (CMF) is the inertial frame in which
$R\equiv 0$, $V\equiv 0$. The velocities of the particles in CMF
are:\ben
(\vec{v}_1){}_{CMF}=g^{-1}_V(\vec{v}_1)=\vec{v}_1-\vec{V}=
{{m_2}\over{M}}(\vec{v}_1-\vec{v}_2)=-{{m_2}\over{M}}\vec{v},\nonumber
\\ (\vec{v}_2){}_{CMF}=g^{-1}_V(\vec{v}_2)=\vec{v}_2-\vec{V}=
{{m_1}\over{M}}(\vec{v}_2-\vec{v}_1)={{m_1}\over{M}}\vec{v},\,\,\,\,\,\,
\la{v12CMF}\een where $\vec{v}=\vec{v}_2-\vec{v}_2$ is the
relative velocity. As a result \ben
(\vec{p}_2)_{CMF}=-(\vec{p}_1)_{CMF}= \hskip 3truecm \nonumber \\
{{m_1m_2}\over{M}}(\vec{v}_2-\vec{v}_1)=
{{m_1m_2}\over{M}}g_{\vec{v}_1}^{-1}(\vec{v}_2)=
-{{m_1m_2}\over{M}}g_{\vec{v}_2}^{-1}(\vec{v}_1). \la{p12CMF}\een
It is natural to define a new effective particle -- {\em the
reduced mass} -- with a mass $m$ and a momentum \ben
\vec{p}:=m\vec{v},\la{p}\een such that \ben
\vec{p}=(\vec{p}_2)_{CMF}=-(\vec{p})_{CMF}.\la{p_p1_p2CMF}\een
Then the comparison of formulas (\ref{p12CMF}) and (\ref{p}) gives
the well known non-relativistic formula for the reduced mass \ben
m={{m_1m_2}\over{M}}.\la{m}\een

\subsection{The Relativistic Two Particle Problem}
There exist several different formal approaches to description of
the relativistic two-particle system. We will illustrate some of
them using the simplest case of two non-interacting particles
$m_1$: $x_1=\{x_1^0, \vec{x}_1\}$, $p_1=m_1u_1$, and $m_2$:
$x_2=\{x_2^0, \vec{x}_2\}$, $p_2=m_2u_2$. Their 4D momenta lie on
the hyperboloids \ben
2\,\varphi_1{}^{free}:=p_1^2+m_1^2\stackrel{w}{=}0,\,\,\,\,
2\,\varphi_2{}^{free}:=p_2^2+m_2^2\stackrel{w}{=}0.
\la{hyperb12}\een

1. One can use the two-times formalism and an action of the
system: \ben {\cal A}_{m_1,m_2}\!=\!-\int\!dt_1\,
m_1\sqrt{1\!-\!\vec{v}_1(t_1){}^2}\!-\int\!dt_2\,
m_2\sqrt{1\!-\!\vec{v}_2(t_2){}^2}. \la{2bA1} \een

2. One can develop a 3D+1 approach, which is not transparently
covariant. Now we are working in the laboratory frame $K$ using
the laboratory time $t=x_1^0=x_2^0$. Then the system is described
by the action \ben {\cal A}_{m_1,m_2}\!=\!-\int\!
dt\left(m_1\sqrt{1\!-\!\vec{v}_1{}^2}\!+\!m_2\sqrt{1\!-\!\vec{v}_2{}^2}\right).
\la{2bA2} \een

3. Finally, one can develop transparently covariant 4D formalism
which is in addition invariant under arbitrary local
re-parameterizations, using the action \ben {\cal
A}_{m_1,m_2}\!=\!\int\!d\tau\left(p_1\dot x_1+p_2\dot x_2-H\right)
\la{2bA3} \een with Hamiltonian \ben
H:={1\over{2\lambda}}(\mu_2\varphi_1{}^{free}+\mu_1\varphi_2{}^{free})\la{H}\een
where $\lambda$ is a Lagrange multiplier the value of which
depends on the choice of the parameter $\tau$, and for the
parameters $\mu_1$, $\mu_2$ we have the relation $\mu_1+\mu_2=1$
and some additional conditions \cite{FT01}.

In any approach the problem is obviously invariant under 4D
translations, described by the group $Tr(4)_{x}$ in the flat 4D
pseudo-Euclidean space-time $\mathbb{E}_x^{(1,3)}$. As a result of
Noether theorem the total momentum \ben
P=p_1+p_2=const\la{totP}\een is a conserved quantity. Using this
quantity we introduce the relativistic CMF $K^U$ as a frame, which
moves relative to the laboratory one with 4D velocity \ben
U=P/\sqrt{-P^2}=P/w\la{U_CMF}\een where $w$ defined by the
equation \ben P^2+w^2\stackrel{w}{=}0\la{w}\een plays the role of
the CM mass. It is easy to check that this quantity has a right
behavior in the non-relativistic limit (NRL):\,\,
$$w\,\stackrel{NRL}{\longrightarrow}\, m_1+m_2=M.$$

Following the same procedure, as in the non-relativistic case, and
using the rules of Lobachevsky geometry for transition to a new
inertial frame -- the relativistic CMF, we obtain the formulas:
\ben (\vec{u}_1){}_{CMF}= \overrightarrow{\Lambda^{-1}_U u_1}=
\hskip 9truecm \nonumber \\
{{m_2}\over{w}}\left((u_2)_0\vec{u}_1-(u_1)_0\vec{u}_2+
{{\left(m_1\vec{u}_1+m_2\vec{u}_2\right)\times
\left(\vec{u}_2\times\vec{u}_1\right)}\over{w+m_1(u_1)_0+m_2(u_2)_0}}\right)
=-{{m_2}\over{w}}\vec{u},\nonumber
\\
(\vec{u}_2){}_{CMF}= \overrightarrow{\Lambda^{-1}_U u_2}= \hskip
9truecm \nonumber \\
{{m_1}\over{w}}\left((u_1)_0\vec{u}_2-(u_2)_0\vec{u}_1+
{{\left(m_1\vec{u}_1+m_2\vec{u}_2\right)\times
\left(\vec{u}_1\times\vec{u}_2\right)}\over{w+m_1(u_1)_0+m_2(u_2)_0}}\right)
=\,\,{{m_1}\over{w}}\vec{u}, \la{u12CMF}\een which are analogous
to the relations (\ref{v12CMF}). Then we see that in the
relativistic case the relations (\ref{p12CMF}) are replaced by the
analogous ones: \ben (\vec{p}_2)_{CMF}=-(\vec{p}_1)_{CMF}=
{{m_1m_2}\over{w}}\,\overrightarrow{\Lambda^{-1}_U u_2}=
-{{m_1m_2}\over{w}}\,\overrightarrow{\Lambda^{-1}_U u_1}=
{{m_1m_2}\over{w}}\,\vec{u}. \la{p12_RelCMF}\een where \ben
\vec{u}=(u_1)_0\vec{u}_2-(u_2)_0\vec{u}_1+
{{\left(m_1\vec{u}_1+m_2\vec{u}_2\right)\times
\left(\vec{u}_1\times\vec{u}_2\right)}\over{w+m_1(u_1)_0+m_2(u_2)_0}}.
\la{uu1u2}\een

Now it is natural to define the new effective particle -- {\em the
relativistic reduced mass} -- with a mass $m$ and a momentum \ben
\vec{p}:=m\vec{u},\la{pRel}\een such that \ben
\vec{p}=(\vec{p}_2)_{CMF}=-(\vec{p})_{CMF}.\la{p_p1_p2RelCMF}\een
Then the comparison of formulas (\ref{p12_RelCMF}) and
(\ref{pRel}) gives the well known non-relativistic formula for the
reduced mass \ben m={{m_1m_2}\over{w}}.\la{mRel}\een

Note that in the non-relativistic case
$g_{\vec{v}_1}^{-1}(\vec{v}_2)=g_{\vec{v}_2}^{-1}(\vec{v}_1)$. In
contrast, in the relativistic case  we have
$\overrightarrow{\Lambda^{-1}_{u_1} u_2}\neq
-\overrightarrow{\Lambda^{-1}_{u_2} u_1}$, due to the nonlinear
character of Lobachevsky velocity space. Nevertheless, we have
$(\Lambda^{-1}_{u_1} u_2)_0= -(\Lambda^{-1}_{u_2}
u_1)_0=(u_1)_0(u_2)_0-\vec{u}_1\cdot\vec{u}_2=u_0$ and this
realation defines $u_0$ component of the velocity of the
relativistic reduced mass. The same result can be obtained
independently from the relations (\ref{hiperb}) and (\ref{uu1u2}):
\ben
u_0=\sqrt{1+\vec{u}\,{}^2}=(u_1)_0(u_2)_0-\vec{u}_1\cdot\vec{u}_2.\la{u0}\een

For components of the momentum of the relativistic reduced mass we
obtain from relations (\ref{hiperb}), (\ref{uu1u2}), (\ref{mRel})
and (\ref{u0}):  \ben  p_0 = { {(p_1)_0(p_2)_0 -
\vec{p}_1\cdot\vec{p}_2} \over {\sqrt{-(p_1+p_2)^2}} } =
-{{p_1\,p_2}\over{\sqrt{-(p_1+p_2)^2}}},\hskip 4.4truecm \nonumber
\\
 \vec{p}={ {(p_1)_0\,\vec{p}_2\!-\!(p_2)_0\,\vec{p}_1} \over
{\sqrt{-(p_1+p_2)^2}}
}+{{(\vec{p}_1\!+\!\vec{p}_2)\times(\vec{p}_1\times\vec{p}_2)}\over
{{\sqrt{\!-(p_1\!+\!p_2)^2}}\left((p_1)_0\!+\!(p_2)_0\!+\!{\sqrt{\!-(p_1\!+\!p_2)^2}}\right)}}.
\la{pRel_compts} \een

In addition we obtain the relations: \ben
(p_1)_0^U=p_0+{{m_1^2}\over{w}},\,\,\,\,\,\,\,\,\,\,\,\,\,\,
(p_2)_0^U=p_0+{{m_2^2}\over{w}};\,\,\,\,\,\,\,\,\, \nonumber \\
(u_1)_0^U={{m_1}\over{w}}+{{m_2}\over{w}}u_0,\,\,\,\,
(u_2)_0^U={{m_2}\over{w}}+{{m_1}\over{w}}u_0;\nonumber\\
\vec{u}_{1,2}\,{}^U =\mp{{m_{1,2}}\over{w}}\, \vec{u}\hskip
4.85truecm \la{p10p20u10u20}\een and \ben
u_0={{w^2-m_1^2-m_2^2}\over {2 m_1 m_2}}, \,\,\,\,p_0={1\over
2}w-{{m_1^2+m_2^2}\over{2 w}},\la{u0_p0}\een \ben
(p_1)_0{}^U={1\over 2}w+{{m_1^2-m_2^2}\over{2 w}},\,\,\,
(p_2)_0{}^U={1\over 2}w+{{m_2^2-m_1^2}\over{2 w}}.\la{p120U}\een
For the parameters $\mu_1$, $\mu_2$ in Eq. (\ref{2bA3}) one
obtains the additional relation \cite{FT01}:\ben
\mu_1-\mu_2={{m_1^2-m_2^2}\over{w^2}}\la{mu1mu2}\een

The key formula is: \ben
\sqrt{\vec{p}_1\,{}^2+m_1^2}+\sqrt{\vec{p}_2\,{}^2+m_2^2}=
\sqrt{\vec{p}\,{}^2+m^2}.\la{sqrt}\een It is valid in the CMF and
allows a reduction of the free two-particle relativistic problem
to the one effective-particle relativistic problem after exclusion
of the CM motion.

It is clear that the formulas for the quantities, related to the
effective particle -- {\em "the relativistic reduced mass"},
reflect the properties of the Lobachevsky geometry in velocity
space, or in the momentum space.

\section{Examples for the Application of the Effective Relativistic Particle}

The general procedure for application of the relativistic reduced
mass in the relativistic two-particles problems in the case of
interacting particles consists in the following steps (See for
example \cite{FT01} and the references therein):

I. In the case of scalar interaction we add the interaction terms
$\Phi_{1,2}$ to the constraint functions $\varphi_{1,2}{}^{free}$
in Eq. (\ref{hyperb12}), thus obtaining
$\varphi_{1,2}:=\varphi_{1,2}{}^{free}+\Phi_{1,2}/2$. Then, using
some general procedure, we exclude the CM motion remaining with
the effective one-particle relativistic Hamiltonian \ben
H:={1\over{2\lambda}}\left(p^2+m^2+\Phi(x)\right). \la{HS}\een

II. In the case of the electromagnetic interactions we use the
standard gauge approach, replacing the components of the 4D
momentum $p_\alpha$ with $p_\alpha -e A_\alpha(x)$.

III. In the case of the gravitational interaction we replace the
pseudo-Euclidean 4D scalar square $p^2$ with the pseudo-Riemannian
one: $p^2=g^{\alpha\beta}(x)p_\alpha p_\beta$, considering the
effective particle as a test particle in an external gravitational
field, described by metric $g^{\alpha\beta}(x)$.

In the general case one can combine the scalar, vector and tensor
interactions, using simultaneously the above procedures.

For illustration let us consider the following examples:

1. In the article \cite{FT01} the effective particle Hamiltonian
\ben H:={1\over{2\lambda}}\left(p_r{}^2+{{J^2}\over{r^2}}+1-
\left(\epsilon+{{e^2}\over{r}}\right)\right)\la{HEM} \een has been
used for the description of relativistic two-particle
electromagnetic interaction. Here $p_r$ is the radial momentum,
$J$ is the total angular momentum, $e^2=-e_1e_2$ is related with
the electric charges $e_{1,2}$, and $\epsilon$ is the
dimensionless effective-particle energy. It was shown that this
Hamiltonian reproduces the right relativistic effects: the orbit,
the perihelion shift, etc. produced by standard approximate
relativistic Hamiltonian up to the terms of order
${{1}\over{c^2}}$.

2. As shown article \cite{FT01} the effective particle Hamiltonian
in Schwarzschild gravitational field: \ben
H:={1\over{2\lambda}}\left(\left(1-\rho{{J}\over{r}}\right)p_r{}^2
+{{J^2}\over{r^2}}+1-
{{\epsilon^2}\over{1-\rho{{J}\over{r}}}}\right)\la{HGR} \een
describes in a correct way the general relativistic two-particles
gravitational interaction up to the terms of order
${{1}\over{c^2}}$. In particular the orbit, the perihelion shift
and a new formula \ben
\omega^*={{\omega^*_S}\over{1-\nu|\varepsilon^*|}}>\omega^*_S\la{omega}
\een for the angular frequency of the last stable orbit (LSO) was
derived. Here $\rho=2m_1m_2G/J$,\,\,
$\nu={{m_1m_2}\over{(m_1+m_2)^2}}$,\,\, $\varepsilon^*=-{1\over
\nu}\left(1-\sqrt{1-2\nu\Big(1-\sqrt{8/9}\Big)}\right)$,\, and
$\omega^*_S$ is the Schwarzschild value of the frequency of LSO.

3. In the article by Crater and Yang (see in \cite{Todorov}) the
effective one-particle approach for the Wheeler-Feynman
action-at-a-distance of the relativistic two-particle system with
scalar and electromagnetic interaction was examined. The final
result is that if one wishes to reproduce the right relativistic
results up to the order ${1\over{c^4}}$,  using the effective
one-particle Hamiltonian, then one has to introduce in this
Hamiltonian the following  modified electromagnetic and scalar
potentials: \ben
 eA_0:={{e_1e_2}\over{r}}+{{5(e_1e_2)^2}(g_1g_2-e_1e_2)\over{12m_1m_2r^3}},
 \nonumber\\
 e\vec{A}:={{5(e_1e_2)^2}(g_1g_2-e_1e_2)\over{12m_1m_2r^3}}\vec{p},\nonumber\\
 \Phi:=-{{g_1g_2}\over{r}}-{{5(g_1g_2)^2}(g_1g_2-e_1e_2)\over{12m_1m_2r^3}}.
 \la{EMmodified}\een Here $g_{1,2}$ are the "scalar charges" of
 the particles.

4. An analogous result was derived by Buonanno and Damour (see in
\cite{Todorov}) for general relativistic two-body problem. They
have introduced an "effective" metric of the form \ben
ds^2{}_{eff}=\left(1+{{a_1}\over{r_{eff}}}+{{a_2}\over{r^2_{eff}}}+
{{a_3}\over{r^3_{eff}}}+\dots\right)dt^2_{eff}-\nonumber \\
\left(1+{{b_1}\over{r_{eff}}}+{{b_2}\over{r^2_{eff}}}+
{{b_3}\over{r^3_{eff}}}+\dots\right)dr^2_{eff}-\nonumber\\
\left(1+{{c_1}\over{r_{eff}}}+{{c_2}\over{r^2_{eff}}}+
{{c_3}\over{r^3_{eff}}}+\dots\right)r^2_{eff} \left(
d\theta^2_{eff}+\sin^2\theta_{eff} d\varphi^2_{eff} \right)
\la{g_eff}\een with a proper coefficients $a_{1,2,3\dots}$,
$b_{1,2,3\dots}$,  $c_{1,2,3\dots}$ to reach the standard 2PN
approximation results.

5. In the very recent articles by Faruque and coauthors (see in
\cite{developments}) the relativistic effective particle approach
was applied to the gravitational interaction of two relativistic
spinning particles using Kerr metric, instead of Schwarzschild
one. The corrections to the orbits and to the perihelion shift:
\ben \delta\varphi = 2\pi\left({1\over{\sqrt{\gamma}}}-1\right)+
{{3\pi}\over{2}}{{\alpha\rho}\over{\sqrt{\gamma}\gamma^2}}\la{kerr}\een
were obtained. Here
$\alpha=1-2\mu\epsilon+3\mu^2\epsilon^2$,\,$\gamma=1-3\beta\mu^2$,
\,$\beta=1-\epsilon$, and  $\mu=a/J$, where $a$ is the specific
spin of the Kerr metric.

\section{Conclusion}
The above examples give a strong indications that the application
of the  method of the relativistic effective one-particle to the
relativistic two particle problem is a powerful tool for the study
of modern topics in physics. This method is based on the
Lobachevsky geometry of the relativistic velocity and momentum
spaces. It deserves further development as a fundamental
theoretical approach to the relativistic problems.

\vskip .3truecm

{\em \bf Acknowledgments} \vskip .truecm  The author is deeply
grateful to the Organizing Committee of the International seminar:
{\em Application and Development of Lobachevsky Ideas in Modern
Physics}, dedicated to 75-th birthday of Professor
N.~A.~Chernikov, Dubna, 25-27 February 2004 for the invitation to
attend this meeting and for the possibility to give this talk.


\begin{thebibliography}{99}\itemsep=0pt
%
\bibitem{Fock} V.~A.~Fock, {\em The Theory of Space, Time and Gravitation},
               Pergamon, Oxford, 1964.

%
\bibitem{LL} L.~D.~Landau, E.~M.~Lifshitz, {\em The Classical Theory of Fields},
        (2d ed.; Reading, Mass: Addison-Wesley, 1962).

%
\bibitem{WF49} J.~A.~Wheeler, R.~P.~Feynman, Rev. Mod. Phys. {\bf
             21}, 425 (1949).
%

\bibitem{K72} E.~H.~Kerner, {\em The Theory of Action-at-a-Distance in
              Realtivistic Particle Dynamics}, Gordon and Breach
              Sci. Publ., NY, 1972.
%
\bibitem{BN} B.~M.~Barbashov, N.~A.~Chernikov, {\em Classical Dynamics of the Relativistic Strings},
             preprint JINR R2-7852, Dubna, 1974 (in Russian);
             B.~M.~Barbashov, V.~V.~Nesterenko, {\em A Model of
             Relativistic String and the Adron Physics},
             Energoatomizdat, Moscow, 1987 (in Russian);
             B.~M.~Barbashov, V.~V.~Nesterenko, {\em Introduction
             to the Relativistic String Theory}, Singapore, World
             Scientific, 1990.
%
\bibitem{ChSh} N.~A.~Chernikov, N.~S.~Shavohina, preprint JINR, Dubna, R2-10375, 1977;
               Theor. Math. Phys. {\bf 42}(1), p. 59 (1980); {\bf 43}(3), p. 356 (1980).
               N.~S.~Shavohina, Izv. Vuzov, Fizika, {\bf 7}, p. 91 (1981);
                                DAN SSSR, {\bf 265}(4), p. 852 (1983);
                                Izv. Vuzov, Fizika, {\bf 7}, p. 66 (1982);
                                Izv. Vuzov, Fizika, {\bf 7}, p. 46 (1983);
                                Problemy teorii gravitacii i
                                elementarnyh chastic, {\bf 15},
                                Energoatomizdat, Moscow, p. 141, 1985;
                                Problemy teorii gravitacii i
                                elementarnyh chastic, {\bf 16},
                                Energoatomizdat, Moscow, p. 189, 1985;
                                Problemy teorii gravitacii i
                                elementarnyh chastic, {\bf 17},
                                Energoatomizdat, Moscow, p. 187, 1986;
                                Mezhdunarodnoe Soveschanie po
                                problemam kvantovoj teorii
                                polia, Dubna, p. 246, 1987. (All references are in Russian.)
%
\bibitem{Todorov} I.~T.~Todorov, Phys. Rev. D{\bf 3}, 2351 (1971).

         I.~T.~Todorov, {\em Quasipotential approach to the
         two-body problem in quantum field theory}, in: Properties of
         Fundamental Interactions at High Energy, ed. A. Zichichi (Ed.
         Compositori, Bologna 1973), 951.
%

         V.~A.~Rizov, I.~T.~Todorov, B.~L.~Aneva, Nucl. Phys.
         B{\bf 98}, 447 (1975).

%
         M.~Malvetti, H.~Pilkuhn, Phys. Rep. {\bf 248}, 1 (1994).

%
         I.~T.~Todorov, {\em Dynamics of relativistic point particles
             as a problem with constraints}, Commun. JINR,
             E-2-10125, Dubna (1976).

%
        A.~Maheshwari, E.~R.~Nissimov, I.~T.~Todorov,
                Lett. Math. Phys. {\bf5}, 359 (1981).

%
        H.~Crater, D.~Yang, {\em A covariant extrapolation of the noninvariant two particle
        Wheeler-Feynman Hamiltonian from Todorov equation and Dirac's constraint
        mechanics}, J. Math. Phys. {\bf 32}, 2374 (1991).
%

        A.~Buonanno, T.~Damour, {\em Effective one-body approach
        to general relativistic two-body dynamics}, gr-qc/981100,
        Phys.Rev. D{\bf 59}, 084006 (1999); {\bf 62}, 064015 (2000).
%
\bibitem{FT01}  Fiziev P.P., Todorov I.T., {\em Effective one-body
approach to the relativistic two-body problem}, E-print:
gr-qc/0010104; Phys. Rev. D{\bf 63}, 104007 (2001).
%

\bibitem{developments}  S. Faruque, {\em Relativistic Periastron Shift of a
Particle in Kerr Field: A Particular Case-Study}, Acta Phys.
Slovaca, (2003).

%
F.De. Polis, G. Ingrosso, A. Nucita, A. Zakhariev,
astro-ph/0310213.

%
T. Damour, Phys. Rev. D{\bf 64}, 124013 (2001), gr-qc/0103018.

%
Faruque~S.~B., Ahsan~M.~A., Ishwar~B., Phys. Lett. A{312}, 166
(2003).

%
Faruque S. B., {\em Circular Orbits of a Particle in Kerr Field: A
study using effective one-body approach} , Acta Phys. Slovaca,
{\bf 53}, 157 (2003).
%
\bibitem{D49} P.~M.~A.~Dirac, Canad. J. Math. {\bf 2}, 129 (1950);
                              Proc. Roy. Soc. A{\bf 246}, 326
                              (1958).

%
        I.~T.~Todorov, {\em Constraint hamiltonian mechanics
        of directly interacting relativistic particles}, in: Relativistic
        Action at a Distance. Classical and Quantum Aspects, ed. J.~Llosa,
        Lecture Notes in Physics {\bf 162} (Springer, Berlin, 1982), 213.





\end{thebibliography}
\end{document}